# Doping Graphene via Organic Solid-Solid Wetting Deposition

Alexander Eberle[a], Andrea Greiner[a], Natalia P. Ivleva[b], Banupriya Arumugam[c], Reinhard Niessner[b], Frank Trixler[a,d],*

[a] *Department of Earth and Environmental Sciences and Center for NanoScience (CeNS), Ludwig-Maximilians-Universität München, Theresienstraße 41, 80333 München, Germany*

[b] *Institute of Hydrochemistry, Chair for Analytical Chemistry, Technische Universität München, Marchioninistr. 17, 81377 München, Germany*

[c] *Institute for Nanoelectronics, Technische Universität München, Theresienstrasse 90, 80333 München, Germany*

[d] *TUM School of Education, Technische Universität München and Deutsches Museum, Museumsinsel 1, 80538 München, Germany*

**Abstract**

Organic Solid-Solid Wetting Deposition (OSWD) enables the fabrication of supramolecular architectures without the need for solubility or vacuum conditions. The technique is based on a process which directly generates two-dimensional monolayers from three-dimensional solid organic powders. Consequently, insoluble organic pigments and semiconductors can be made to induce monolayer self-assembly on substrate surfaces, such as graphene and carbon nanotubes, under ambient conditions. The above factuality hence opens up the potential of the OSWD for bandgap engineering applications within the context of carbon based nanoelectronics. However, the doping of graphene via OSWD has not yet been verified, primarily owing to the fact that the classical OSWD preparation procedures do not allow for the analysis via Raman spectroscopy – one of the main techniques to determine graphene doping. Hence, here we describe a novel approach to induce OSWD on graphene leading to samples suitable for Raman spectroscopy. The analysis reveals peak shifts within the Raman spectrum of graphene, which are characteristics for p-type doping. Additional evidence for chemical doping is found via Scanning Tunneling Spectroscopy. The results open up a very easily applicable, low-cost, and eco-friendly way for doping graphene via commercially available organic pigments.

* Corresponding author. Tel.: +49 89 2179 509. E-mail: trixler@lrz.uni-muenchen.de

## 1. Introduction

Nearly a decade ago, it became evident that the miniaturisation of silicon-based electronics is limited and that it will soon reach its termination [1-2]. As a result, numerous scientists began exploring the prospects of carbon-based nanoelectronics, to utilise the outstanding electronic properties of graphene, and thus to increase the performance of existing and to develop future electronic devices like the flexible or inkjet-printed electronics [2-10]. However, a deeper insight revealed the production of functional nano-systems, like the nanoscale transistors, to be quite a challenging task [10-14]. A promising approach nonetheless, for the fabrication of essential semiconductive sub-regions, came forward as providing a bandgap to the graphene substrate, by covering it with a monolayer of an organic semiconductor [6-8]. Such a monolayer, in turn, can be built up by the bottom-up technologies (as the mostly available the vapor deposition- [15] or the liquid phase deposition- [16] techniques), directing the self-assembly of organic molecules via the non-covalent interactions (hydrogen bonding, Van-der-Waals, π–π stacking, and electrostatics) [10-14].

However, the processing of organic semiconductor and pigment molecules imposes its own limitations: only few of these compounds survive the thermally enforced vacuum sublimation unscathed that is necessary to apply vapour deposition methods, as the organic molecular beam deposition technique [16-18]. Further, as most of the organic pigments with promising semiconductive properties are insoluble in common liquids, liquid phase deposition techniques like the drop-casting or spin-coating [18] call for an additional chemical functionalisation [18-21]. Nevertheless, in relation to the standard pigments employed usually in the industrial sector, the custom synthesis of functionalised semiconductors marks as an extensive and cost-intensive process.

As an alternative approach, we hence developed the Organic Solid-Solid Wetting Deposition (OSWD) technology, an environmental friendly, cheap, up-scalable, and both, straightforward and quick to perform procedure [19-22]. The OSWD being based on the solid-solid wetting effect [23-25], it is the gradient of surface free energy that acts as the prime driving force behind the technology. Briefly summarising its basics, it can be said that the surface free energy of organic semiconductor crystals that physically contact an inorganic substrate like graphite, graphene, carbon nanotubes or $MoS_2$ [20], gets modified when appropriate organic or aqueous dispersing agents are used. As a consequence, a solid-solid wetting process is triggered, detaching semiconductor molecules from the attached crystal and adsorbing them to the substrate surface. Subsequently, the adsorbed molecules assemble into supramolecular architectures, covering the substrate surface [19-21].

However, in this regard, worth investigating was, if the OSWD generated surface coverage dopes the graphene substrate and thus induces a bandgap. Since a bandgap alters the spectroscopic properties of a material, such a modification, if induced, can be detected via the Raman spectroscopic analysis [6-8]. Raman spectroscopy has been known as a fast and a non-destructive high-resolution technique, which can be employed to study the fundamental physical properties of carbon nanomaterials, such as determining their layer thickness, detecting structural defects, and verifying graphene doping [6,34-39]. It can thus be said to be a reliable and widely used method for investigating the doping of graphene. However, in order to perform an accurate Raman measurement, the substrate surface needs to be covered with adsorbate layers, freely accessible for the laser beam and featuring homogeneity of the order of few hundred nanometres magnitude. Unfortunately, the hitherto used standard OSWD preparation technique fails to generate such a covering, thereby calling for a modification of the approach.

Hence, in an attempt to modify the OSWD technique for gaining samples suitable for the Raman spectroscopy analysis, a series of experimental tests were performed, their results being presented and discussed in the following sub-sections. In this regard, initial efforts were made to enhance the surface coverage of the sample substrate by incorporating a reworking step. Furthermore, a new 'thermally triggered' sample preparation technique was tested and is put forward, with an aim of potentially triggering the OSWD process without employing a catalysing dispersing agent. Post successful generation of suitable Raman samples, the results of the Raman spectroscopy analyses are discussed, as to determine whether the OSWD produced supramolecular surface covering modifies the substrate's electronic properties by providing a bandgap or not. Finally, the results of a series of Scanning Tunneling Spectroscopy (STS) tests are presented, as to verify the outcomes of the Raman analysis via an independent, additional experimental technique. STS is especially suited in this regard, owing to its sensitivity in probing the chemical doping of graphene, by providing an atomic resolution analysis of the local electronic properties of a surface [26-32].

## 2. Results and discussions

*2.1. Refinement of the standard sample preparation method by incorporating a reworking step*

As per the results of the previous investigations, the substrate surface coverage generated via the OSWD can be altered to a large extent by substituting the dispersing agent in use, i.e. without replacing the organic semiconductor itself [20]. Until now, for the samples fabricated via the standard preparation method, the maximum achievable surface coverage rate was limited to approx. 67 %, not being sufficient enough for the execution of Raman spectroscopy measurements. Thus, to accomplish such analysis and to enhance the surface coverage of the 'traditionally prepared' samples, the incorporation of a reworking procedure was considered. For this, and to explore the physio-chemical basics of the OSWD process, model systems out of highly oriented pyrolytic graphite (HOPG) as the substrate material, and the organic semiconductive pigment gamma quinacridone (γQAC) as the active phase were utilized. γQAC is known to be a cheap and commercially available pigment with promising electrical properties, low toxicity, an excellent physical and chemical stability, and with biocompatibility for applications in the living organism [40-46]. Furthermore, the gamma polymorph has been known to be the most stable out of the four possible, three-dimensional crystal structures, this polymorph being built up by the linear QAC molecules (refer Fig. 1 (a)) connected with their neighbours via four hydrogen bonds of the type NH···O=C [47]. Presuming the successful processing of three-dimensional γQAC crystals into substrate surface adsorbate structures through the OSWD approach, the quinacridone molecules (QAC) have been investigated to arrange themselves in one-dimensional supramolecular chains via the NH···O=C hydrogen bonds. Further, multiple parallel and side-by-side appearing chains have been reported to form supramolecular arrays (refer Fig. 1 (b)) [19-20]. Note: the abbreviation QAC is used for quinacridone in general, usually relating to either quinacridone molecules or quinacridone adsorbate structures, whereas the term γQAC is employed for the 3D gamma polymorph of quinacridone.

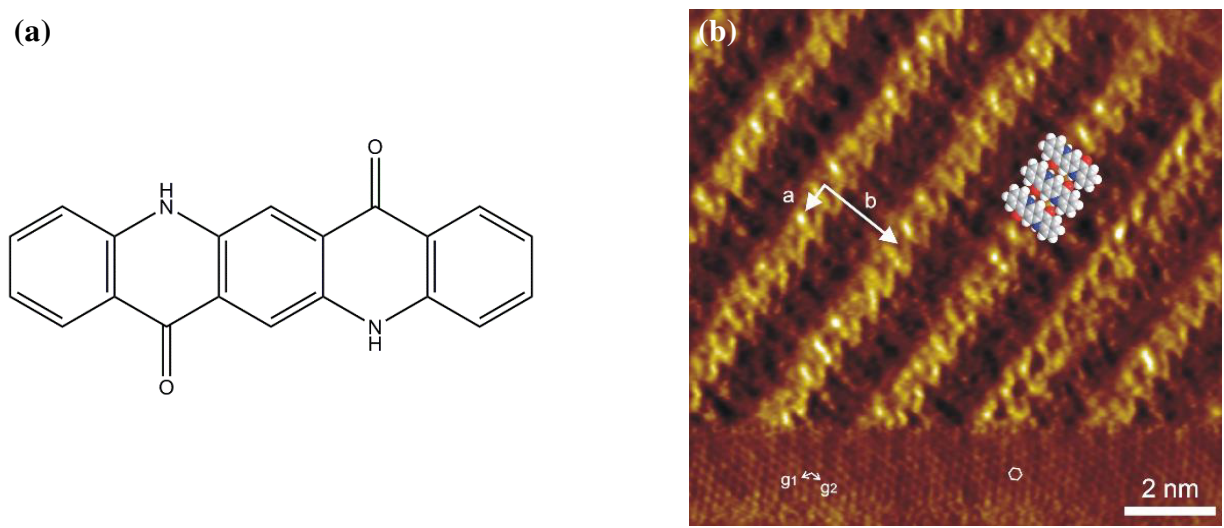

**Fig. 1.** OSWD induced monolayer self-assembly of QAC on graphene. (a) Chemical structure of the QAC molecule. (b) Upper section: supramolecular QAC structures situated atop single layer graphene on copper as the substrate, with the observed structures having lattice

parameters as: a = 0.72 ± 0.02 nm, b = 2.06 ± 0.02 nm, and an intermediate angle of 89 ± 2 °. Further, the superimposed inlay on the QAC structures (little right of the image) depicts the arrangement of single QAC molecules in one-dimensional chain-like structures. The bottom of the image, in addition, presents the underlying graphene substrate's structure, with lattice parameters of the graphene unit cell as: g1 = g2 = 0.246 nm. Besides, the white marked hexagon in the image (atop the graphene substrate) represents one carbon ring of the graphene structure, with an atom to atom distance of 0.142 nm.

Force field calculations in this respect revealed that a γQAC crystal comprises of at least one crystal face, in which the QAC molecules have binding energy less than the binding energy of a molecule adsorbed on a graphene substrate [19]. From these calculations, it can be hence deduced that only on the condition that the γQAC crystal contacts the HOPG with one of its energetically favourable crystal faces, QAC molecules can detach and subsequently attach themselves to the HOPG substrate and thus initiate the self-assembly processes. In addition, experiments revealed that a complete coverage of the HOPG surface by the supramolecular QAC arrays could not be achieved, although the standard sample preparation technique covers the entire HOPG surface with a distinct layer of γQAC crystals (i.e. γQAC powder) (refer Fig. 2 (a)). From the above theoretical and experimental results, it can hence be deduced that the OSWD approach is an anisotropic process. The latter deduction, in turn, proposes a way of subsequently increasing the surface coverage of the sample, by gently rubbing the remaining γQAC powder against it. Such a procedure, supposedly, forces the γQAC crystals to roll over the HOPG surface, thereby significantly increasing the chances of specific crystal faces to contact the HOPG.

For the execution of the above, the virgin HOPGs were hence initially treated with a dispersion of γQAC and the dispersing agent octylcyanobiphenyl (8CB), the latter is known to be one of the few dispersing agents that neither does vaporise at room temperatures nor disturbs the STM measurements (further information on the 8CB's chemical structure and its ability to self-assemble stable and well-ordered arrays being available in the supplementary data). Subsequent STM measurements of the samples prepared in the above manner revealed an overall surface coverage of 50 ± 4 %, including twice the standard deviation (refer Fig. 2 (a)). Thereinafter, using a metal spatula, the remaining γQAC powder was gently rubbed against the substrate, and consequently the results depict a greater overall surface coverage of 98 ± 2 % (refer example picture in Fig. 2 (b)). Hence, it can be said that the incorporated reworking step enabled almost complete surface coverage, though the surface covering displayed still various arrays with different orientations. Therefore, since the substrate surface covering obtained by the above sample preparation procedure does not exhibit homogeneity of the order of magnitude of a few hundred nanometres, the above sample preparation method cannot be hence used to prepare samples suited for accurate Raman spectroscopy measurements.

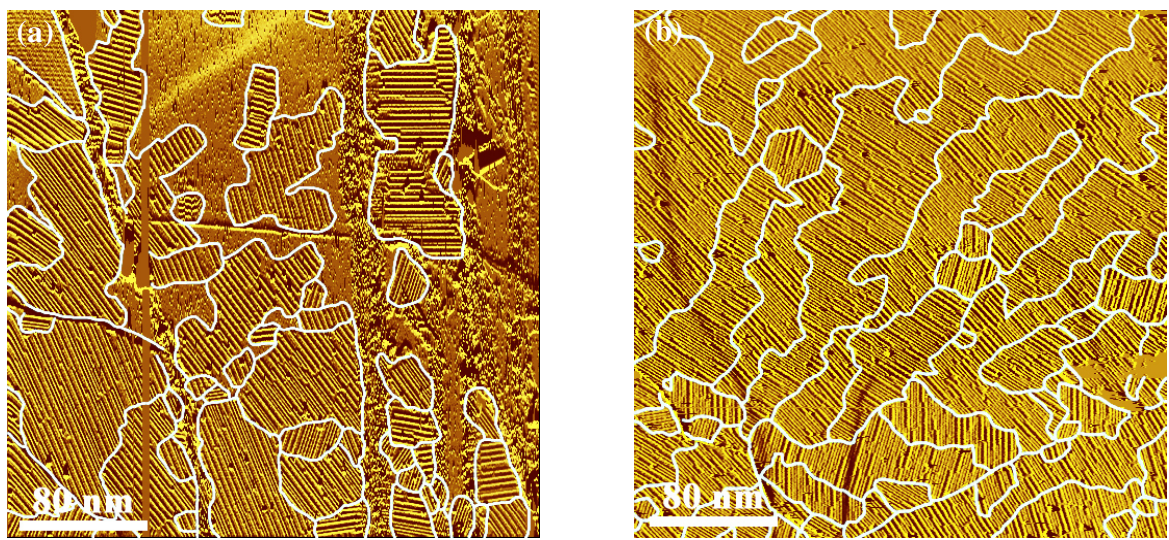

**Fig. 2.** Large-area STM scans of supramolecular QAC arrays (white markings highlighting the borders of the QAC arrays). (a) Example STM image of a supramolecular surface covering, atop a HOPG substrate, generated using a dispersion of γQAC and 8CB as the dispersing agent; the average surface coverage rate being 50 ± 4 %. (b) Example STM image post gently rubbing the remaining γQAC powder with 8CB being still present; the average surface coverage in this case being 98 ± 2 %.

## 2.2. Triggering the OSWD without a catalysing dispersing agent

Since all attempts to rework samples fabricated by the standard OSWD sample preparation method did not lead to sufficient Raman samples, it was hence thought upon to develop a new and adequate sample preparation technique that could supply and transfer the essential activation energy to trigger the OSWD in an alternative way, i.e. without the need of a catalysing dispersing agent. A way of doing so, as per literature, could be by employing the concept that the gradient of surface free energy at the solid-solid interface changes with an increase in temperature [50-51], thereby presenting the possibility of triggering the solid-solid wetting effect via a thermal sample treatment [23-25]. However, for implication of such a treatment, the thermal stability and the melting point of the involved pigment has to be taken into account. Since QAC crystals are thermally stable up to their melting point of 390 °C [46,52], a series of tests were performed where the virgin HOPGs covered with pure γQAC powder were heated up gently to different temperatures, as to trigger the OSWD process thermally. Results in this respect revealed, though generation of no supramolecular QAC structures for temperatures up to 160 °C, however, the detection of a significant surface coverage of 83 ± 13 % for samples being further heated up to 240 °C (refer Fig. 3). In addition, it was observed that the supramolecular QAC chains arranged themselves in a large-scale homogeneous monolayer, which changed its orientation almost exclusively by hitting the border to a new graphite plane. Such planes, in turn, are predetermined by the quality of the substrate surface. Further experiments were performed in this regard, where several HOPGs covered with γQAC powder were heated up to approx. 270 °C, yielding a greater surface coverage of 92 ± 6 %. Hence, it can be concluded that the above described thermally triggered sample preparation method marks as a promising approach towards fabricating samples, enabling accurate Raman spectroscopy measurements.

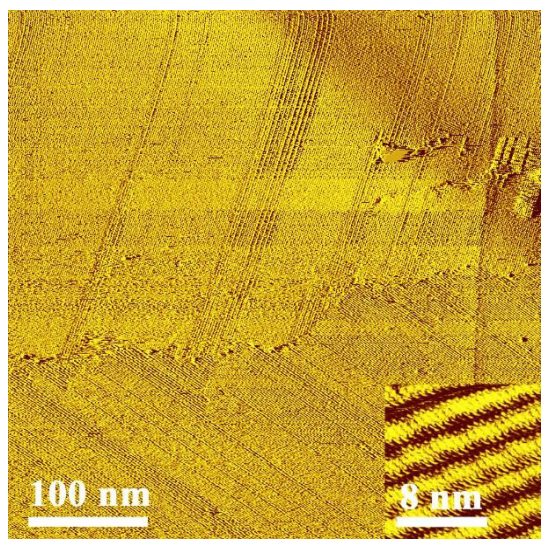

**Fig. 3.** STM image of a HOPG substrate covered with a QAC monolayer. The sample was prepared by heating up dry γQAC powder on the HOPG substrate to a temperature of 240 °C; the surface coverage rate being 83 ± 13 % in this case. The close-up view in the bottom right corner highlights, how the QAC molecules arrange themselves within the adsorbate layer.

In order to determine the underlying formation mechanism that results in the observed extended and well-ordered QAC adsorbate layers, worth recapitulating, initially, are the so far gained findings about the classical OSWD sample preparation. Summarizing briefly, when a three-dimensional semiconductor crystal contacts a HOPG substrate, molecules from the crystal detach and get attached to the substrate, provided the adsorption energy $E_a$ is higher than the cohesive energy $E_c$ of the semiconductor bulk (as derived from force field calculations in previous studies [19]). In this respect, having analysed HOPG samples via STM measurements over a period of several weeks (with semiconductor particles and 8CB being continuously present), no significant increase in the array dimensions could be detected over time. Further, except few results showing instable bilayer structures, no case of three-dimensional growth could be detected [20]. In this regard, an approximate energetic criterion was found to predict two-dimensional vs. three-dimensional growth under conditions of thermodynamic equilibrium [53-54]: the condition for the three-dimensional growth being $E_c < E_a$ and the inverse being true for the two-dimensional growth. Hence, it can be said that for the HOPG and γQAC model system as used in the present study, growth of two-dimensional morphology is expected to be favoured, what corresponds to our findings.

Generally speaking, the growth of a supramolecular surface adsorbate structures is proportional to the surface diffusion. Surface diffusion was observed for both single adsorbate molecules and compact adsorbate clusters containing numerous molecules [54], provided the diffusion barrier is overcome. Further, the surface diffusion process is thermally promoted, just as in the case of bulk diffusion, with diffusion rates (corresponding to the adsorbate mobility) increasing with increasing temperature. We can thus conclude from our experimental findings that the surface diffusion is limited in the temperature range of up to 160 °C, whereas significant diffusion is achieved for temperatures of 240 °C and above. Hence, provided the conditions for a thermally triggered OSWD prevail, the adsorbed QAC molecules migrate over the substrate surface in a direction away from the depositing γQAC crystal plane, with the concentration gradient and diffusion processes as the driving forces. As a result, further molecules can be deposited from the γQAC source, leading to an expansion of the QAC array.

As the formed adsorbate layers show a high degree of order, an additional thermal annealing effect is presumed. In this respect, it is referred to experiments conducted by Wagner et al., analysing the thermal annealing of a two-dimensional surface covering formed by QAC arrays exhibiting different orientations, however using Ag(111) as the substrate material [46]. Their results revealed that for a temperature range between 550 – 570 K (i.e. 277 – 297 °C), the structural properties of the QAC covering change towards the formation of extended and well-ordered monolayers. Hence, in analogy with the above, it can be said that for our study the thermal annealing presumably plays a part, by triggering the rearrangement of QAC molecules and thus leading to the formation of extended and highly ordered monolayers. In addition, worth mentioning are the series of continuative experiments performed to test the stability of these thermally generated QAC adsorbate structures [20]. Our results yielded no signs of structural decomposition after the samples have been stored for 36 days under ambient conditions and furthermore, the QAC adsorbates were observed to be resistant towards humidity and direct water contact.

Discussing the effect of sintering processes on the above, it can be said that first of all, the working temperatures (240 °C and 270 °C respectively) are way low than the melting temperature of γQAC (390 °C) [52]. Although early reports on lower temperature sintering observed with nanoscale particles conjectured a melting temperature reduction, this idea however has been dispelled by careful analysis, revealing further no new mechanism to be active in sintering nanoscale particles beyond known processes [55]. In this respect, as per the well-known viscous flow sintering model, the concept of sintering is analogous with the growth of sinter necks between contacting objects (i.e. grains in this context), connecting the contacting grains and forming a polycrystalline solid [55]. Hence, it can be said that a potential sintering would both interlink contacting semiconductor particles to strongly bonded crystalline structures and connect these structures to the substrate surface at the contact points. Further, in contrast to the OSWD process, the resulting sinter neck formation would be isotropic in nature, resulting in numerous contacting points, thereby establishing a permanent connection (besides, the type of chemical bonding between the substrate and the semiconductor is supposed to be π–π stacking). Thus, small-scale nanocrystals fixed to the substrate could be detected directly via STM, whereas the presence of large, permanently fixed nanocrystals would be noticed since they would considerably disturb the STM measurements, thereby making STM imaging hardly possible. However, STM measurements revealed no detection of sintered γQAC crystals of any kind, thereby highly limiting the influence of sintering processes on the above thermally triggered sample treatment approach.

## 2.3. Replacing QAC by DMQAC

Successfully applied to a HOPG substrate, the newly developed, thermally triggered sample preparation method generated a surface covering that synced with the requirements of a Raman spectroscopic analysis. Testing the applicability of OSWD for single graphene layer on a copper foil as the substrate material (refer Fig. 1) revealed similar supramolecular structures as detected on a HOPG substrate [20]. Hence, the single QAC molecules arranged themselves in one-dimensional chain-like structures, leading to the coverage of the substrate surface by multiple parallel and side-by-side appearing chain-like formations. Further, the lattice parameters 'a' and 'b' of the supramolecular monolayer (compare Fig. 1) were observed to correspond to one of the array configurations as observed on the HOPG substrates, such an HOPG array configuration being termed as the 'relaxed QAC chain configuration' in this case [20].

Next, the initial Raman measurements of samples fabricated via thermally triggered OSWD, applied to graphene as the substrate material, revealed that due to the sole presence of $sp^2$-bonded carbon atoms in both the γQAC and the graphene substrates, the location, the shape, and the intensity of the corresponding G peaks (described in the next sub-sections) was found to be quite similar for both the materials. Consequently, the spectra of both the samples could not be distinguished accurately, making hence the further, exact analysis quite a challenging task. Thus, to resolve the above, it was decided to replace γQAC by the quinacridone derivate dimethylquinacridone (DMQAC). In this regard, the linear DMQAC molecules (refer Fig. 4) generate three-dimensional crystal structures, iso-structural to the $α^I$ polymorph formed by the QAC molecules [47-49].

In this respect, to begin with, different samples were analyzed via STM measurements, to explore the processability of DMQAC via the OSWD technique. Results revealed, analogous to QAC, the DMQAC molecules arranging themselves in one-dimensional supramolecular chains, forming in turn two-dimensional arrays. Further, within the limits of accuracy, the lattice parameters of these supramolecular structures were found to be identical for both the substrates HOPG and single layer graphene (refer Fig. 5). Besides, the latter substrate depict the distinct honeycomb structure of single layer graphene [26] (refer Fig. 5 (b)). However, further analysis revealed that in contrast to the QAC adsorbate structures (refer Fig. 3) [20], the DMQAC chains arrange themselves solely in a close-packing chain configuration (on both the HOPG and the single layer graphene), leading to DMQAC arrays with high packing density (refer Fig. 5 and Fig. 6).

In addition, regarding the adsorbate layer thickness, most of the observed DMQAC arrays were clearly determined to be two-dimensional monolayers (refer Fig. 5 and Fig. 9 (b)). However, previous studies using DMQAC revealed that, although rarely observed, bilayer and even trilayer structures could as well be detected, with their structure being similar to the ones seen when analysing QAC adsorbates [20]. Further, these structures were found to range in the size of single DMQAC chains up to arrays of a few dozen nanometres in diameter. Hence, for clarity purposes, the term 'DMQAC adsorbate layer' will be used hereinafter. Nevertheless, it should be noted that since the Raman spectroscopy is an averaging technique, the related analysis of potential doping effects is not disturbed by sporadically occurring small-sized bilayer or trilayer adsorbates. Furthermore, as extended multilayer adsorbates would generate fluorescence effects within the Raman signal, their occurrence could be detected, however,

...

Raman analysis revealed no detection of multilayer DMQAC adsorbates (refer to the Raman discussion below).

Results revealed that the thermally triggered sample preparation method yielded an overall surface coverage of 92 ± 8 % atop a HOPG substrate, with the DMQAC adsorbate layer found to be sufficiently homogeneous in nature (refer Fig. 6). However, worth noting here is that the structure of the copper foil leads to a rather uneven surface, making hence large-scale STM scans of the covered graphene samples impossible. Thus, we have not been able to determine the surface coverage of the graphene samples accurately, though, the promising results of the HOPG samples and the explored similar adsorbate structure properties on both the substrates indicated similar coverage rates. In addition, the STM analysis of graphene samples over a period of four weeks detected negligible decomposition of the DMQAC adsorbate structures, thus indicating the temporal stability and the resistance against humidity of the latter adsorbate layers being similar to that of QAC adsorbates [20].

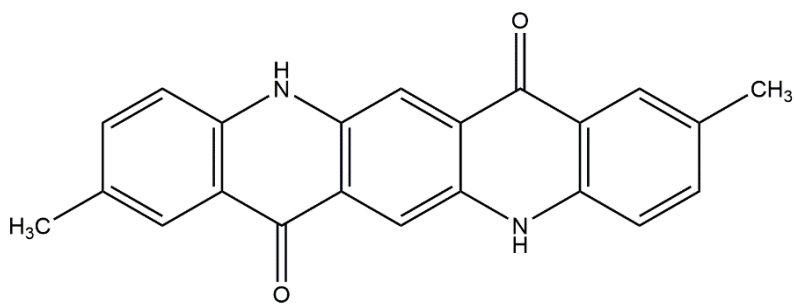

**Fig. 4.** Chemical structure of the DMQAC molecule.

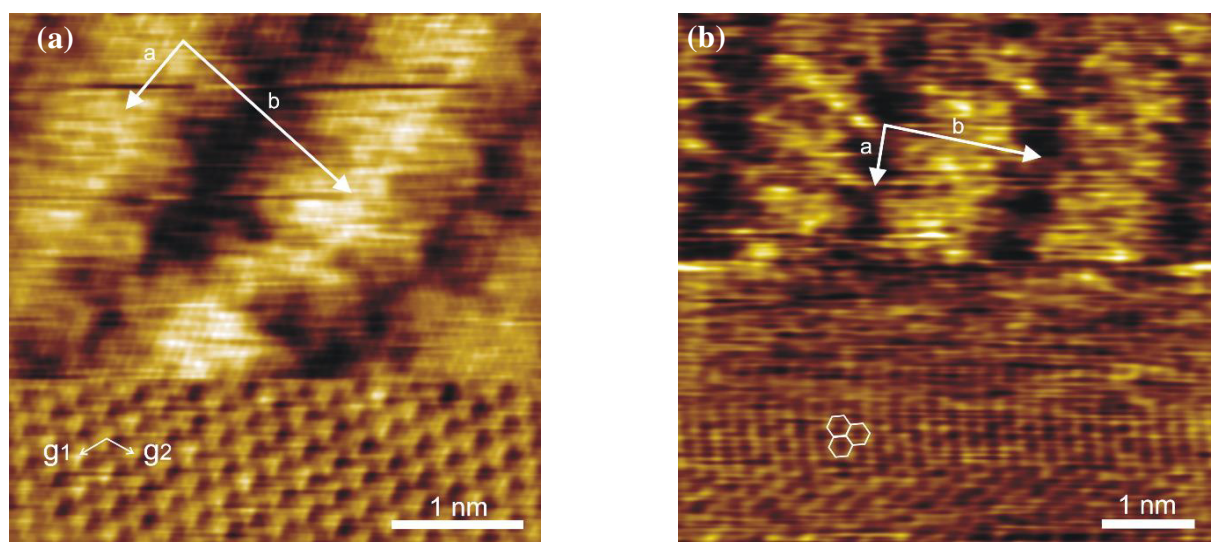

**Fig. 5.** STM images of supramolecular DMQAC structures (upper sections) and the subjacent substrates (lower section). The STM pictures have been equalized using the substrates unit cell parameters for calibration. (a) The substrate in use is HOPG and the lattice parameters of the adsorbate structures were found to be as: a = 0.68 ± 0.02 nm, b = 1.72 ± 0.02 nm, and an intermediate angle of 87 ± 2°. The HOPG unit cell is depicted via the parameters g1 and g2, having lengths as: g1 = g2 = 0.246 nm. (b) Using single layer graphene on a copper foil as the substrate, the lattice parameters of the adsorbates were determined to be as: a = 0.67 ± 0.02 nm, b = 1.72 ± 0.02 nm, and an intermediate angle of 88 ± 2°. Further, the marked hexagon in the image represents one carbon ring of the graphene structure; the atom to atom distance in this regard being 0.142 nm.

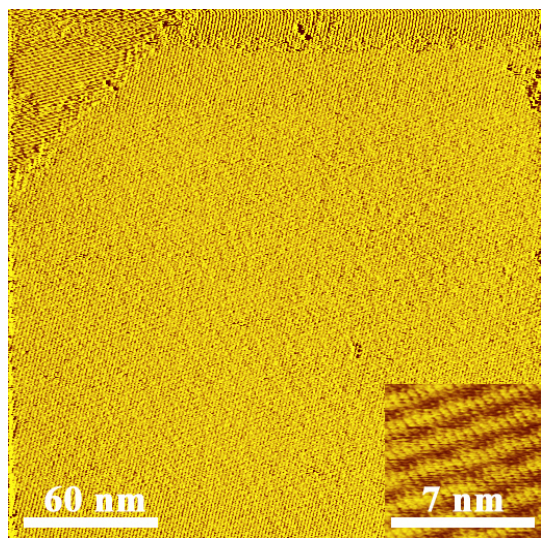

**Fig. 6.** Example STM picture of a two-dimensional supramolecular adsorbate layer atop a HOPG substrate generated by DMQAC molecules, yielding an overall average coverage rate of 92 ± 8 %. Besides, the close-up inset in the bottom right corner highlights, how the DMQAC molecules arrange themselves within the adsorbate layer.

*2.4. Raman spectroscopy measurements*

Generally speaking, the Raman spectrum of carbon-based substrates can be mainly characterised by three characteristic peaks [6,34-39], i.e. the D peak, the G peak, and the 2D peak. The D peak is typically observed at a Raman frequency of approx. 1350 cm$^{-1}$ indicating a structural defect, owing to its activation due to $A_{1g}$ mode breathing vibrations of six-membered sp$^2$ carbon rings, which are absent in defect-free graphene [34-35]. Hence, the D peak intensity increases with the amount of disorder present in the material [36]. The G peak, on the other hand, appears at approx. 1580 cm$^{-1}$, being associated with the doubly degenerate $E_{2g}$ phonon at the Brillouin-zone centre [34-35]. Finally, the 2D peak is the second order of the D peak, found usually at about 2680 cm$^{-1}$. Further, since the 2D peak originates from a process where momentum conservation is satisfied by two phonons with opposite wave vectors, it is always present for graphene, with its activation requiring no structural defects [34-36].

Moreover, variations of the above characteristic Raman peaks can be generated by the introduction of either mechanical strain or by chemical doping. However, both these sources cause specific changes in the Raman spectrum, making hence the respective variations distinguishable from each other [35-36]. Mechanical strain, for example caused by an adsorbate layer or by a previous thermal treatment of the sample, modifies the crystal phonons due to changes in the lattice constants and the resulting structural disorder activates the D peak [36]. It was further found that the compressive strain leads to an upshift of the G and the 2D peak, the tensile stress whereas, leading to the downshift of both these peaks. However, in either case, the 2D peak shift is several times greater, with the intensity ratio of the 2D to G peak ($I_{2D} / I_G$) remaining unaltered [35-36]. Nevertheless, in contrast to the above, the intensity ratio $I_{2D} / I_G$ has been observed to be sensitive to chemical doping. Appropriate doping effects cause the above ratio to decrease monotonically with an increase in both the electron and the hole concentration [6,36-38]. Also, as per empirical findings, doping with electron-donating aromatic molecules (i.e. electron- or n-type doping) downshifts the G peak frequency, whereas the presence of electron-withdrawing molecules (i.e. hole- or p-type doping) leads to a G peak upshift. Nonetheless, the 2D frequency is reported to be upshifted, irrespective of the type of doping [37-38].

The results of the Raman test series are as presented in the Fig. 7. To begin with, the depicted Raman spectra were determined by averaging nine measurements, both for the pure graphene and the 'DMQAC powder on a graphene substrate' samples, and by averaging 16 measurements for the 'graphene covered with a DMQAC adsorbate layer' sample. Results revealed detection of no graphene-specific peaks at the appropriate peak locations (Fig. 7 (a)) within the spectrum of the DMQAC powder on a graphene substrate (treated with the identical, thermally triggered sample preparation method). However, a few DMQAC-specific peaks could be observed which could be determined precisely in the spectra of both the DMQAC samples (DMQAC powder and DMQAC adsorbate layer on graphene), their locations being determined as $1204 \pm 2$ cm$^{-1}$, $1233 \pm 2$ cm$^{-1}$, $1312 \pm 1$ cm$^{-1}$, and $1567 \pm 2$ cm$^{-1}$, respectively. Also, further analysis in this regard revealed no shift of the DMQAC-specific peaks in the spectra of both the DMQAC samples. Furthermore, worth noting here is that the spectrum of the sample 'graphene covered with a DMQAC adsorbate layer' exhibited additional peaks, which are as well related to DMQAC [56], their determined locations being as $1408 \pm 1$ cm$^{-1}$, $1509 \pm 1$ cm$^{-1}$, and $1648 \pm 2$ cm$^{-1}$, respectively. Nevertheless, the precise location of these peaks could not be determined in the spectrum of the sample 'graphene covered with DMQAC powder', due to the occurrence of fluorescence (Fig. 7 (b)) [57-58]. In this regard, the revealed finding that fluorescence effects are quenched by the properties and conditions of the sample 'graphene covered with a DMQAC adsorbate layer', indicate though a structural transition of the three-dimensional DMQAC particles into a thin adsorbate layer (being triggered by the OSWD process), thereby acting no longer as a bulk solid [59-60]. This deduction affirms the findings of the previously presented STM measurements.

Further investigation revealed that the spectra of the pure graphene substrate and that of the graphene sample covered with a two-dimensional DMQAC adsorbate layer comprises the graphene-specific peaks, i.e. the D, the G, and the 2D peaks (Fig. 7 (a)). The location of the D peak was found to be similar for both the samples (Fig. 7 (b)), however, owing to its significantly lower intensity in contrast to the other peaks, it being determined only at around $1348 \pm 5$ cm$^{-1}$ (including twice the standard deviation). In addition, the D peak intensity was as well found to be similar for both the samples, thereby indicating the exclusion of thermally induced structural defects. Furthermore, for the pure graphene substrate, the G peak was found at $1592 \pm 1$ cm$^{-1}$ and the 2D peak at $2691 \pm 1$ cm$^{-1}$, respectively, whereas for the 'graphene plus DMQAC adsorbate layer' sample, their locations being $1595 \pm 2$ cm$^{-1}$ and $2701 \pm 3$ cm$^{-1}$, respectively (Fig. 7 (c) and (d)). Thus, the Raman spectrum of the DMQAC adsorbate layer sample revealed an upshift of both the G peak frequency ($3 \pm 2$ cm$^{-1}$) and the 2D peak frequency ($10 \pm 3$ cm$^{-1}$), indicating thereby chemical doping with electron-withdrawing aromatic molecules (i.e. p-type doping). The latter is further supported by literature findings, showing that DMQAC thin films act only as p-type materials [30,33]. Comparison of the Raman intensity ratio $I_{2D} / I_G$ of the pure graphene substrate ($I_{2D} / I_G = 1.15 \pm 0.06$) and the graphene substrate covered with a DMQAC adsorbate layer ($I_{2D} / I_G = 0.53 \pm 0.06$) further revealed a significant decrease, indicating hence the chemical doping of graphene as well. In addition, mechanical strain was excluded as the probable source of the Raman peak shifts, since no increase of the D peak intensity was found after the thermally triggered sample preparation and due to the decrease in the Raman intensity ratio $I_{2D} / I_G$.

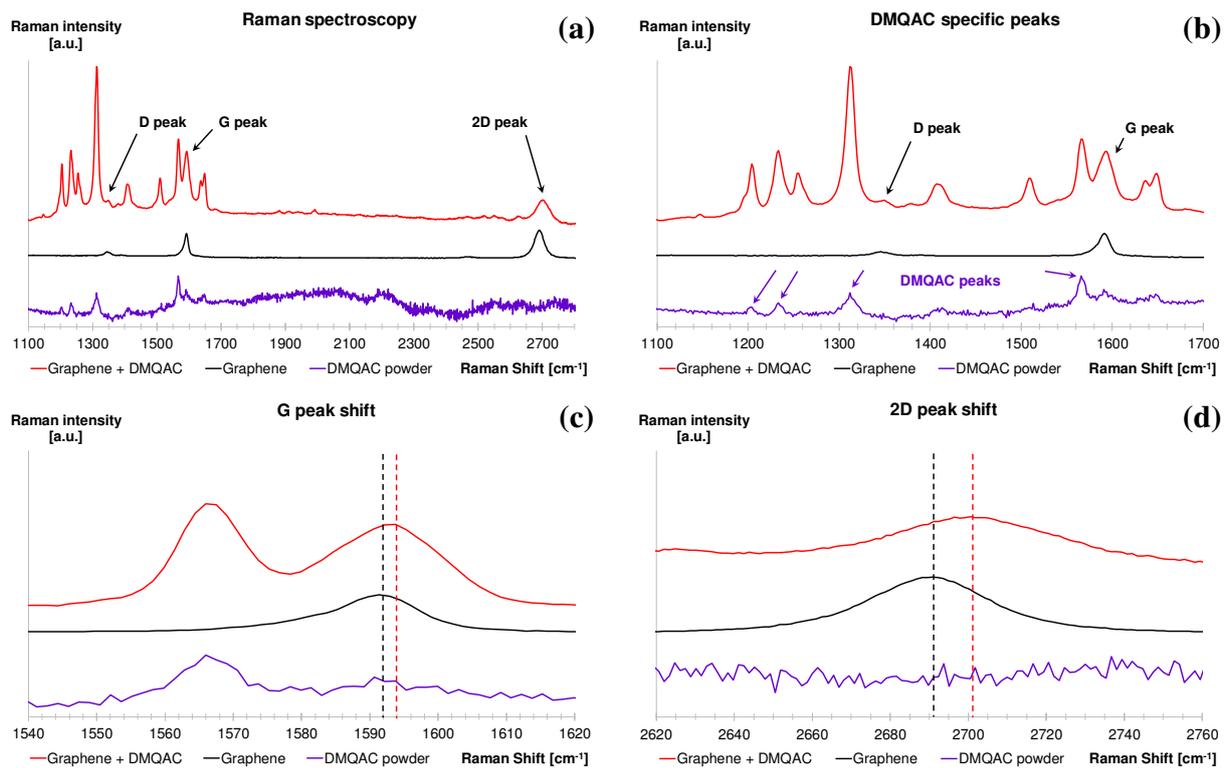

**Fig. 7.** Data representing the averaged results of the performed Raman measurements. (a) Overview of the Raman spectra in the relevant frequency range between 1100 and 2800 cm$^{-1}$. (b) Zoomed-in view of the frequency range between 1100 and 1700 cm$^{-1}$. The Raman spectra of both the DMQAC powder and the two-dimensional DMQAC adsorbate layer atop a graphene substrate depict additional peaks. Further, a close-up view of the frequency range revealing: (c) the G peak and (d) the 2D peak, respectively.

## 2.5. Scanning Tunneling Spectroscopy (STS) analysis of DMQAC on graphene

In order to further verify the results of Raman spectroscopy by an independent experimental technique, an STS test series was conducted in addition. STS, determining the current-bias spectra I(V) at a fixed tip position, is known as a sensitive technique to probe the local electronic properties of a surface [26-32]. At low tip–sample voltages, the tunneling differential conductance is proportional to the local density of states of conducting and semiconducting samples [26-29]. However, due to the dependence of sample-tip separation on the tunneling probability, the STS acquisition relies on the initial set point tunneling conditions [28]. Thus, the tunneling distance for the below discussed STS spectra was adjusted with identical tunneling parameters, whenever possible: bias = 50.1 mV, and tunnel current = 1 nA for the analysis of graphene, and bias = 1.5 V, and tunnel current = 501 pA for the analysis of both DMQAC arrays and the test series regarding potential chemical doping. Further, each spectrum was acquired within 100 ms.

Appropriate samples were investigated via STS measurements, in order to explore the surface electronic structure of graphene substrates with DMQAC adsorbates atop. To begin with, STS spectra taken of pure graphene are as shown in the Fig. 8 (a). As can be seen, the curve progressions are in agreement with the reported metallic behaviour of the zero-gap semiconductor graphene [27]. Further, the spectra taken of DMQAC arrays feature a sample bias range with approximately zero current (refer Fig. 8 (b)), as expected of a semiconducting material [27]. However, the curves are subject to strong fluctuations that are related to the ambient measurement conditions, causing thermal fluctuations that affect the STS measurement accuracy [27]. Thereby, owing to the above, the direct determination of the tunneling differential conductance was hardly possible. Instead, a trend line of the type $f(x) = a(x+b)^3$ (with 'a' and 'b' as constants) was fitted to the obtained spectra and differentiation yielded suitable parabolic shaped dI/dV curves (refer Fig. 8 (b) and Fig. 9 (a)). Assuming a differential conductance below 0.7 nA to be zero, the bandgap of DMQAC was estimated to be 2.4 ± 0.2 eV (including twice the standard deviation), the result hence being in good accordance with the reported HOMO – LUMO gap of DMQAC of 2.3 eV [30].

Regarding the analysis of a potential chemical doping, it was reported that a suitable semiconducting surface adsorbate modifies the electronic properties of the substrate beyond the physical borders of the adsorbate. Hence, it was found that the STS measurements yield a decreasing bandgap in the direction away from the chemical dopant [31-32]. The results of the related test series are shown in the Fig. 9. As can be seen, the determined band gap decreases almost linearly in the direction away from the DMQAC array, reaching the detection limit of 0.1 ± 0.2 eV at a distance of 8 nm. The latter result thus signalizes the chemical doping of graphene via supramolecular DMQAC adsorbates, generated via the OSWD.

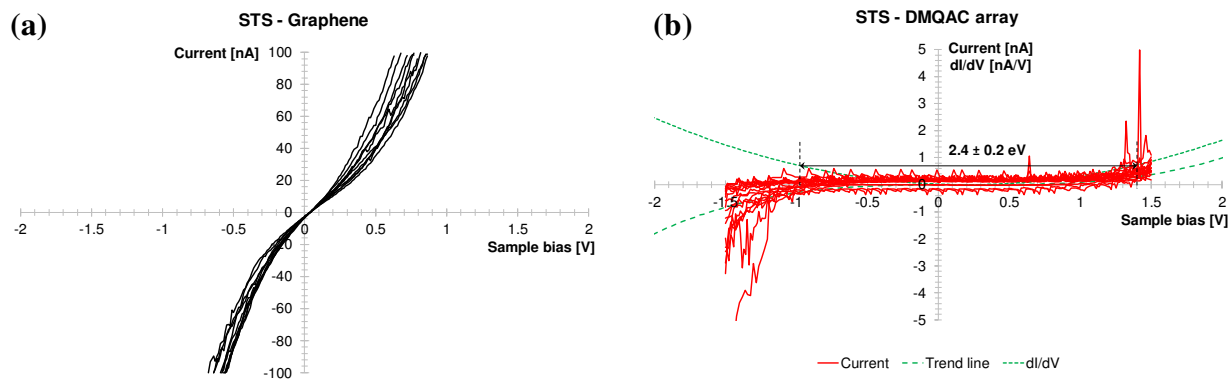

**Fig. 8.** STS measurements at different locations of three different (but equally prepared) samples. (a) 10 STS measurements of pure graphene. (b) 18 STS measurements of supramolecular DMQAC adsorbates. Further, the indicated trend line and the dedicated dI/dV curve exemplarly reveal the determination of a bandgap (for further details, refer to the explanations in the text); the corresponding bandgap of a DMQAC monolayer being found as 2.4 ± 0.2 eV (including twice the standard deviation).

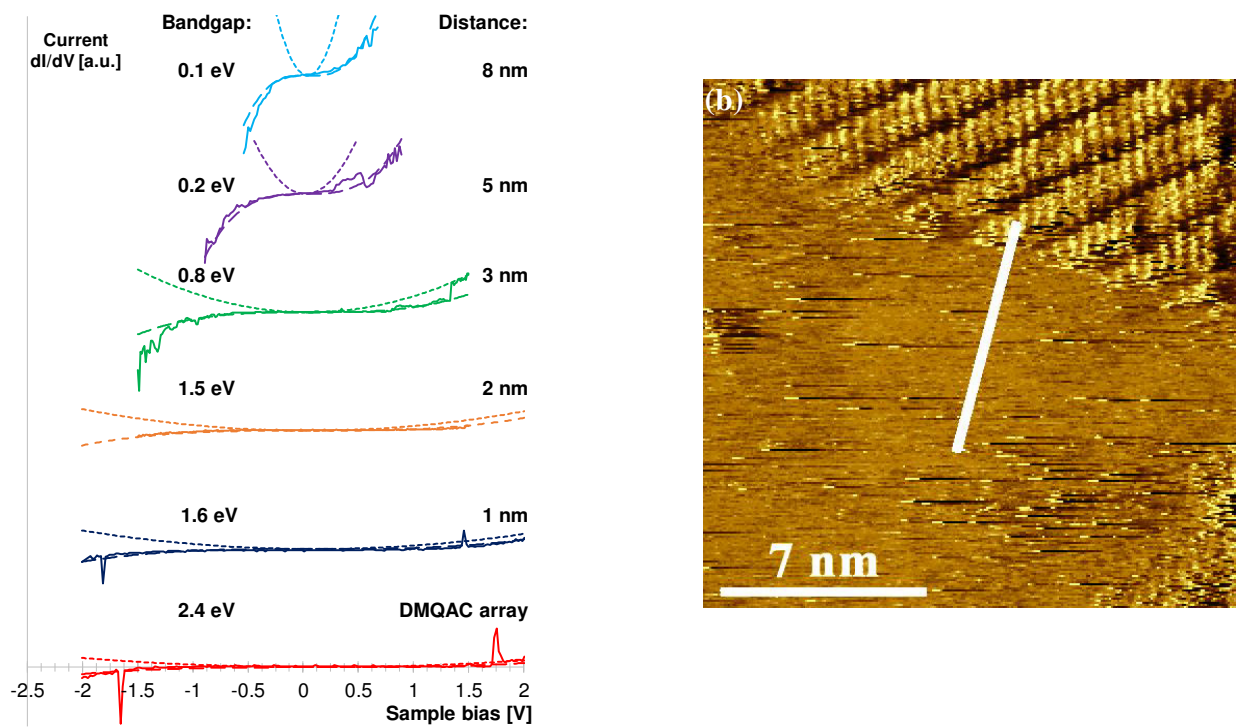

**Fig. 9.** (a) Example STS measurements, taken at different distances from a DMQAC array (including the calculated trend lines and the dI/dV curves). With the determined bandgaps indicated above the related spectrum, twice the standard deviation was deduced to be ± 0.2 eV.

(b) STM image of the region, at which the STS measurements were taken. The upper section depicts a DMQAC monolayer and the lower section shows the bare graphene substrate. The measurement points were located on the highlighted white line.

## 3. Conclusions

Our approach to thermally trigger the solid-solid wetting of crystalline carbon surfaces by organic semiconductor particles enables an easily applicable technique to achieve monolayers with high surface coverage rates and uniform adsorbate structures. Moreover, graphene samples generated in this way allows for the analysis of possible doping effects via techniques as Raman spectroscopy. Additionally, by using the commercially available pigment DMQAC for the new OSWD approach, clear spectral evidence of chemical doping effects of graphene could be obtained. This finding is further supported by STS analysis, showing evidence of chemical doping by DMQA adsorbate structures. The results hence bring forward new and straightforward to perform approaches for the fabrication and bandgap engineering of low-cost, but large-scale products based on pigment-functionalized graphene, like the printed and potentially flexible carbon based electronics [9].

**Acknowledgements**

The authors would like to thank Michael Blum for his kind support with the STM measurements using graphene as the substrate material. Heartfelt acknowledgment as well to Michaela Wenner for her assistance in exploring the potentials of sample reworking. We also would like to extend our gratitude to Arthur Wesemann for his contribution with experiments and valuable experiences, in regard to the imaging of organic semiconductor adsorbates via the NaioSTM setup (as a precondition for our STS studies). Also, we would like to thank Neeti Phatak for her support with the proofreading and editing of this publication. The Bayerisches Staatsministerium für Umwelt und Verbraucherschutz is gratefully acknowledged for their funding.

**Appendix A. Supplementary data**

Supplementary data related to this article can be found at xxx.

*A. Eberle et al. / Carbon 00 (2017) 000–000* 22[34] M.W. Iqbal, A.K. Singh, M.Z. Iqbal, J. Eom, Raman fingerprint of doping due to metal adsorbates on graphene, J. Phys. Condens. Matter. 24 (2012) 335301. doi:10.1088/0953-8984/24/33/335301.

[35] W.X. Wang, S.H. Liang, T. Yu, D.H. Li, Y.B. Li, X.F. Han, The study of interaction between graphene and metals by Raman spectroscopy, J. Appl. Phys. 109 (2011) 15–18. doi:10.1063/1.3536670.

[36] C. Casiraghi, Probing disorder and charged impurities in graphene by Raman spectroscopy, Phys. Status Solidi - Rapid Res. Lett. 3 (2009) 175–177. doi:10.1002/pssr.200903135.

[37] X. Dong, D. Fu, W. Fang, Y. Shi, P. Chen, L.J. Li, Doping single-layer graphene with aromatic molecules, Small. 5 (2009) 1422–1426. doi:10.1002/smll.200801711.

[38] H. Liu, Y. Liu, D. Zhu, Chemical doping of graphene, J. Mater. Chem. 21 (2011) 3335. doi:10.1039/c0jm02922j.

[39] Z. Ni, Y. Wang, T. Yu, Z. Shen, Raman Spectroscopy and Imaging of Graphene, Nano Res. 1 (2008) 273–291. doi:10.1007/s12274-008-8036-1.

[40] E.D. Głowacki, R.R. Tangorra, H. Coskun, D. Farka, A. Operamolla, Y. Kanbur et al., Bioconjugation of hydrogen-bonded organic semiconductors with functional proteins, J. Mater. Chem. C. (2015) 6554–6564. doi:10.1039/C5TC00556F.

[41] E.D. Głowacki, G. Romanazzi, C. Yumusak, H. Coskun, U. Monkowius, G. Voss et al., Epindolidiones-Versatile and Stable Hydrogen-Bonded Pigments for Organic Field-Effect Transistors and Light-Emitting Diodes, Adv. Funct. Mater. (2014) 1–12. doi:10.1002/adfm.201402539.

[42] M. Sytnyk, E. D. Głowacki, S. Yakunin, G. Voss, W. Schöfberger, D. Kriegner et al., Hydrogen-Bonded Organic Semiconductor Micro- And Nanocrystals: From Colloidal Syntheses to (Opto-)Electronic Devices, J. Am. Chem. Soc. 136 (2014) 16522–16532.

[43] E.D. Głowacki, M. Irimia-Vladu, M. Kaltenbrunner, J. Gsiorowski, M.S. White, U. Monkowius et al., Hydrogen-bonded semiconducting pigments for air-stable field-effect transistors, Adv. Mater. 25 (2013) 1563–1569. doi:10.1002/adma.201204039.

[44] E.D. Glowacki, L. Leonat, G. Voss, M. Bodea, Z. Bozkurt, M. Irimia-Vladu et al., Natural and nature-inspired semiconductors for organic electronics, Org. Semicond. Sensors Bioelectron. IV. 8118 (2011) 81180M–81180M–10. doi:10.1117/12.892467.

Supporting Information

# Doping Graphene via Organic Solid-Solid Wetting Deposition

*Alexander Eberle[a], Andrea Greiner[a], Natalia P. Ivleva[b], Banupriya Arumugam[c], Reinhard Niessner[b], Frank Trixler[a,d],\**

[a] Department für Geo- und Umweltwissenschaften and Center for NanoScience (CeNS), Ludwig-Maximilians-Universität München, Theresienstraße 41, 80333 München, Germany

[b] Institute of Hydrochemistry, Chair for Analytical Chemistry, Technische Universität München, Marchioninistr. 17, 81377, München, Germany

[c] Institute for Nanoelectronics, Technische Universität München, Theresienstrasse 90, 80333 München, Germany

[d] TUM School of Education, Technische Universität München and Deutsches Museum, Museumsinsel 1, 80538 München, Germany

---

\* Corresponding author. Tel.: +49 89 2179 509; fax: +49 89 2179 239.
*E-mail address:* trixler@lrz.uni-muenchen.de (F. Trixler)

## Materials and methods

**Sample preparation**

As a standard Organic Solid-Solid Wetting Deposition (OSWD) sample for scanning tunnelling microscope (STM) investigations, a dispersion with 2 wt% of the pigment γQAC (5,12-Dihydro-quino[2,3-b]acridine-7,14-dione, purchased as Hostaperm Red E5B02 from Clariant) dispersed in 4 ml of the dispersing agent 8CB (purchased as 4'-n-Octylbiphenyl-4-carbonitrile from Alfa Aesar, item no. 52709-84-9) was prepared. A few drops of this dispersion were then dispensed on a highly ordered pyrolytic graphite (HOPG, supplier NT-MDT, item no. GRBS/1.0), to trigger the OSWD at the interface between the dispersed pigment particles and the HOPG. Besides, single layer graphene on a copper foil (suppliers: Graphene Laboratories, item no. CVD-Cu-2X2, and Graphenea Inc.) was used as the substrate material for further tests. Also, as an alternative sample preparation method (to thermally trigger the OSWD), the HOPG substrate was fully covered with the powdered pigment, but without a catalysing dispersing agent. The covered substrate was then heated up to 240 °C and 270 °C, respectively, using a special hotplate, enabling an accurate temperature control and providing a smooth temperature increase (Stuart SD160, temperature accuracy ± 1.0 °C). In any case, the ready-made STM samples were investigated within days, and as per the previous tests, QAC arrays were observed unaltered in their structure for a minimum of four weeks, unless not influenced via any external forces [1].

Further, for the Raman experiments, single layer graphene on Si/SiO$_2$ (purchased from Graphene Laboratories, item no. 1ML-SIO2-5P) was used as the substrate material and DMQAC (2,9-Dimethyl-5,12-dihydro-quino[2,3-b]acridine-7,14-dione, purchased as Hostaperm Pink E from Clariant) as the organic semiconductor. In order to induce thermally triggered OSWD on the substrate graphene/Si/SiO$_2$, a small amount of powdered DMQAC, enough to cover the substrate surface, was added on top of the substrate. The substrate (along with the powdered organic semiconductor atop) was then placed on the previously mentioned hot plate, heating the sample up to approx. 270 °C. Once the sample was heated well for the given temperature, it was taken from the hot plate and the pigment powder was immediately removed from the substrate surface by mechanical shaking. Being still hot, the pigment powder does not adsorb humidity from the surrounding environment and thus does not stick to the substrate surface. Hence, the appropriately prepared Raman samples in the above manner enabled an accurate Raman spectroscopy analysis, being free of measurements artefacts related to traces of bulk pigment particles. Moreover, before performing any Raman measurements, the Raman sample was chilled to room temperature under ambient conditions.

**STM and STS settings**

Two types of STM systems operating under ambient conditions were used for this study. First was a home-built STM combined with a SPM 100 control system, supplied by RHK Technology Inc.; the scans settings being: bias = 1 V, tunnel current = 300 pA, and the line time = 50 ms. Secondly, a commercial STM, type NaioSTM, supplied by Nanosurf GmbH, was employed for the measurements, as depicted by fig. 5, 8 and 9 within the main article; the scans settings for imaging DMQAC being: bias = 1.5 V, current = 501 pA, and line time = 80 ms, and for imaging graphene being: bias = 50.1 mV, tunnel current = 1 nA, and line time = 60 ms. In addition, the voltage pulses used to improve the scan quality were set in the range between 4.3 and 10 V.

The STS measurements were performed using the NaioSTM. The tunneling distance was adjusted with tunneling parameters being similar to the ones as mentioned above. Each spectrum was acquired within 100 ms; both the STM and the STS measurements being performed under ambient conditions. All STS measurements were performed at randomly distributed positions and from 3 different (but equally prepared) samples, to exclude incidental findings. Further, the graphene samples used for the STS measurements were prepared via the standard OSWD sample preparation method, thereby using a dispersion of DMQAC and 8CB. This preparation method was preferred owing to two reasons: First, STM and STS measurements require a conductive sample, thus excluding graphene on SiO2/Si and making the use of single layer graphene on copper foil mandatory. However, according to the specifications of the manufacturer Graphenea Inc., single layer graphene on the copper foil is only thermally stable up to 60 °C. Tests further in this regard, applying the thermally triggered OSWD sample preparation method to single layer graphene on a copper foil, revealed significant damage of the substrate, thereby making accurate STM and STS measurements impossible for graphene/Cu samples prepared via thermally triggered OSWD. Secondly, OSWD induced by the dispersing agent 8CB enables to achieve STS under controlled ambient conditions without undefined contamination layers. STS experiments with pure 8CB on graphene/Cu yielded no evidence of chemical doping by 8CB adsorbate layers (refer example STS measurements in Fig. 1).

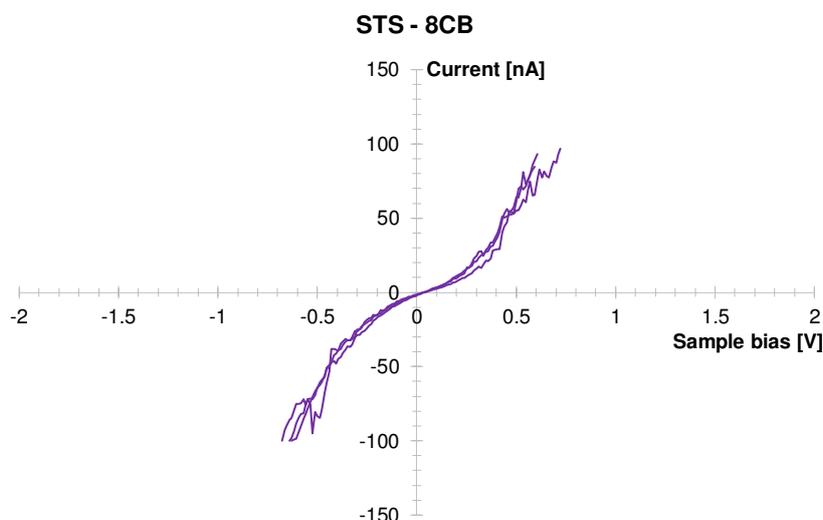

**Fig. 1.** Example STS measurements of 8CB adsorbate layers atop single layer graphene.

**Image processing**

For analysing the supramolecular adsorbate structures in the STM images, the software SPIP™ (Scanning Probe Image Processor, Version 2.3000; distributor: Image Metrology A/S) was used. Image distortions by the drift of a STM scan were analysed and corrected by applying two-dimensional Fast Fourier Transformation (FFT) to the images, done by using the known lattice parameters of the substrates as a reference. Autocorrelations of corrected images were employed for measuring the distances and angles in the substrates and the adsorbates. To minimize noise in the final STM images, a selective FFT filtering was applied with thresholding between 15-25 (min.) and 100 (max.).

**Determining the surface coverage**

To determine the coverage of the HOPG surface by the QAC arrays within a single STM picture, the software Gwyddion (64bit), version 2.42 was used. For this, initially, the QAC arrays via the tool "Mask Editor" were highlighted, followed by the export of single array dimensions by the "Grain distributions" tool, this finally being accompanied by the Microsoft Excel 2013 calculations to determine the coverage ratio. Further, to investigate the average coverage of a STM sample, per sample an area of about 0.7 µm² was analysed. This was done by using a number of STM pictures with high scan resolution and without measurement artefacts, the images were further randomly selected from at least five clearly separated positions on the covered substrate; the average coverage rates, including the double standard deviations, being specified in the current publication.

**Raman Spectroscopy**

The Raman experiments were performed using a LabRAM HR Evolution Raman System, provided by HORIBA Scientific and controlled via the software LabSpec 6. Further, the tests were performed using a frequency-doubled Nd:YAG laser, having a wave length of 532 nm and applying a laser power output of 0.84 mW on the sample. Besides, a diffraction grating with 600 lines per mm and a confocal pinhole with 100 mm diameter were employed. The wavelength calibration was realized by focusing the laser on a silicon wafer and analysing the first order phonon band of silicon at 520 cm$^{-1}$. Since the DMQAC powder sample shows no first order phonon band of silicon at 520 cm$^{-1}$, the wavelength calibration for this sample was done using the DMQAC peak at 1312 cm$^{-1}$. Furthermore, the intensity correction algorithm of the LabSpec 6 software was used to adjust the Raman intensity variations caused by the Raman measurement system. To compensate for the occurrence of a strong fluorescence effect while analysing the 'graphene plus DMQAC powder' samples, the measurements were adjusted via a baseline correction, by applying a polynomial of the sixth degree.

# Additional information

**8CB**

With respect to the result presented in the present publication, it should be noted that 8CB is known to self-assemble stable and well-ordered arrays that can be detected via STM [2] (for the chemical structure of 8CB refer Fig. 3). Having used 8CB in numerous STM experiments, it was found that the 8CB arrays and arrays built by semiconductor molecules can sometimes be imaged at the same time using identical STM scan settings, and sometimes not. In this regard, we propose that the difficulties in imaging both the adsorbate structures simultaneously are related to the orientation of the liquid crystal 8CB with respect to the substrate surface. STM detectable 8CB adsorbate structures form only when the liquid crystal is oriented in such a way that the molecules are aligned parallel to the substrate. So, although no 8CB can be seen in the figures depicted in relation to the reworking experiments, 8CB arrays could however be found on the HOPG surface (refer Fig. 2). Further, we could never find bilayer or multiple layer structures built by alternating layers of QAC and 8CB arrays to date. Thus, it is assumed that the uncovered areas around QAC arrays most likely contain 8CB arrays, although they sometimes cannot be imaged via STM. In addition, results indicate that the QAC arrays exhibit a stronger affinity to the HOPG surface than the 8CB arrays. This is probably attributed to a strong π-π interaction between the fully condensed aromatic ring system of the QAC molecules and the graphene substrate, whereas the 8CB molecule providing just two phenyl groups enabling a π-π interaction and a weakly interacting alkyl chain (for the chemical structure of the 8CB molecule refer to the supporting information). Hence, it is assumed that the QAC molecules compete successfully for array formation, and furthermore that 8CB arrays are expelled by the expanding QAC arrays. The latter assumption is additionally supported by the finding that the reworked HOPG surface almost exclusively contains supramolecular QAC structures, whereas 8CB arrays being hardly found on such a sample.

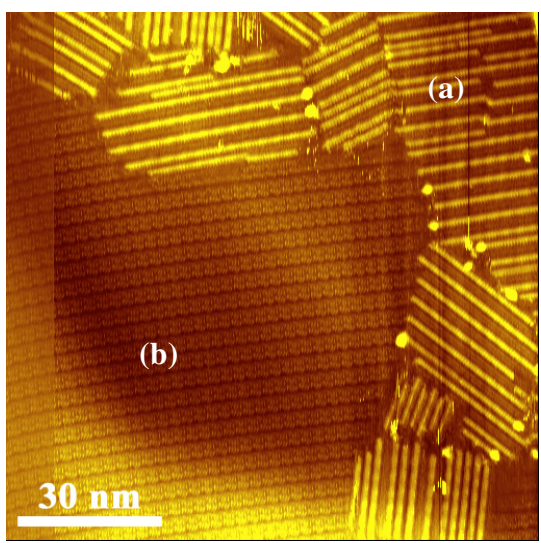

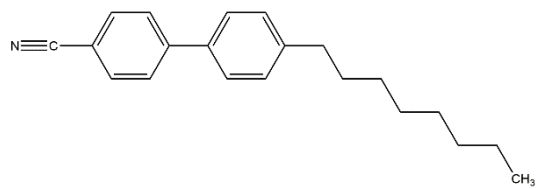

**Fig. 2.** STM image of a HOPG substrate treated with a dispersion of γQAC and 8CB. (a) Supramolecular QAC array. (b) Array formed by 8CB molecules.

**Fig. 3.** Chemical structure of the 8CB molecule.

**References**

[1]  A. Eberle, A. Nosek, J. Büttner, T. Markert, F. Trixler, Growing low-dimensional supramolecular crystals directly from 3D particles, CrystEngComm. (2017). doi:10.1039/C6CE02348G.

[2]	T.J. McMaster, H. Carr, M.J. Miles, P. Cairns, V.J. Morris, Adsorption of liquid crystals imaged using scanning tunneling microscopy, J. Vac. Sci. Technol. A Vacuum, Surfaces, Film. 8 (1990) 672–674. doi:10.1116/1.576370.